\documentclass[twocolumn,english,pre,showpacs]{revtex4}
\usepackage[T1]{fontenc}
\usepackage[latin9]{inputenc}
\usepackage{amsmath}
\usepackage{amssymb}
\usepackage{graphicx}
\usepackage{esint}
\usepackage{amsfonts}
\usepackage{array}
\usepackage{multirow}
\usepackage{amstext}
\usepackage{CJK}
\usepackage{babel}
\usepackage{babel}
\usepackage{babel}
\usepackage{subfigure}

\begin{document}

\begin{CJK}{GBK}{song}

\title{The Thermodynamical Instability Induced by Pressure Ionization in
Fluid Helium}

\author{Qiong Li$^{1}$, Hai-Feng Liu$^{1}$, Gong-Mu Zhang$^{1}$, Yan-Hong Zhao$^{1}$, \\ Guo Lu$^{1}$, Ming-Feng Tian$^{1}$, Hai-Feng Song$^{1}$}
\affiliation{$^{1}$Laboratory of Computational Physics, Institute of
Applied Physics and Computational Mathematics, \\ Beijing 100094,
China }

\begin{abstract}
A systematic study of pressure ionization is carried out in the
chemical picture by the example of fluid helium. By comparing the
variants of the chemical model, it is demonstrated that the behavior
of pressure ionization depends on the construction of the free
energy function. In the chemical model with the Coulomb free energy
described by the Pad\'e interpolation formula, thermodynamical
instability induced by pressure ionization is found to be manifested
by a discontinuous drop or a continuous fall and rise along the
pressure-density curve as well as the pressure-temperature curve,
which is very much like the first order liquid-liquid phase
transition of fluid hydrogen from the first principles simulations.
In contrast, in the variant chemical model with the Coulomb free
energy term empirically weakened, no thermodynamical instability is
induced when pressure ionization occurs, and the resulting equation
of state achieves good agreement with the first principles
simulations of fluid helium.
\end{abstract}


\maketitle

\section{Introduction}

The thermodynamical properties of hydrogen/deuterium/helium have
been intensively studied \cite{Ceperley2012} over the past decades,
due to its applications in astrophysics \cite{SCvH1995} and in
inertial confinement fusion (ICF) research
\cite{Hu2010PRL,Morales2012,Wang2013}. Over the wide range of
densities and temperatures, the regime of partial
dissociation/ionization has attracted particular attention, with the
focus on the description of pressure dissociation/ionization, which
is also of fundamental interest in condensed matter physics and
remains a controversial problem. For helium, the studies
\cite{Ebeling1990CPP,FKE1992,SCvH1995,WC2005,JSR2003} in the
framework of chemical models found a first order transition
typically known as plasma phase transition; nonetheless, the
first-principles simulations \cite{Kietzmann2007,Militzer2009}
reached the opposite conclusion based on the calculated
pressure-density relations which are perfectly smooth without any
indication of first order transition.

In contrast, for hydrogen/deuterium, the first order transition
connected with pressure ionization/dissociation was reported by both
the chemical model \cite{Beule1999PRB} and the first-principles
simulations
\cite{Scandolo2003,Bonev2004b,Vorberger2007,LHR2010,Morales2010b}.
For instance, Ref.\cite{Scandolo2003} found a first order
liquid-liquid transition for compressed hydrogen, which is signaled
by large density fluctuations in a constant-pressure ensemble. By
inspecting the pressure isochores,
Refs.\cite{Bonev2004b,Vorberger2007} also reported signatures of a
first order liquid-liquid transition, by showing a discontinuous
drop in pressure on increasing temperature for deuterium
\cite{Bonev2004b}, or by showing a smooth pressure isochore with a
region of negative slope for hydrogen \cite{Vorberger2007}. By
inspecting the pressure isotherms for dense liquid hydrogen,
Refs.\cite{LHR2010,Morales2010b} further predicted the clear
evidence of a first order liquid-liquid transition, which is a large
drop in pressure on increasing density \cite{LHR2010} or a smooth
pressure isotherm with a plateau \cite{Morales2010b}. Note that
these first-principles simulations, although in agreement about the
existence of the liquid-liquid phase transition, yield different
results on the transition pressure and temperature
\cite{Ceperley2012}. Besides, there are also a few numerical studies
such as Ref.\cite{Delaney2006}, which do not report the first order
liquid-liquid transition.

Obviously, whether pressure dissociation/ionization occurs smoothly
or via a first order phase transition remains an open question. In
partially ionized plasma of helium, there is no dissociation
equilibrium interfering with the ionization equilibrium. Hence,
helium is a better candidate than hydrogen for studying pressure
ionization. In this paper, we are devoted to a close exploration of
pressure ionization by the example of fluid helium, in the framework
of chemical models similar to those used in
Refs.\cite{SCvH1995,JSR2003,Schwarz2005,Chen2007}, and try to
resolve the contradiction between the chemical models
\cite{Ebeling1990CPP,FKE1992,SCvH1995,WC2005,JSR2003} and the first
principles simulations \cite{Kietzmann2007,Militzer2009} about the
prediction of plasma phase transition in fluid helium, by examing
the construction of the free energy function.

The rest of the paper is organized as follows. In Sec. II, the model
is described. In Sec. III, the results and discussions are
presented. Finally, the conclusions and outlooks are given.

\section{The model}

\subsection{The chemical picture}

The chemical picture assumes the existence of distinguishable
chemical species - molecules, atoms, ions and electrons, which are
interacting and reacting in equilibrium. For partially ionized
plasma of helium, the chemical species include He, $\text{He}^{+}$,
$\text{He}^{2+}$ and $e$, and the plasma composition is determined
by ionization equilibrium equations, which are derived by minimizing
the free energy density with respect to the abundances of ionic
species. The well-known Saha equation is established for
non-interacting classical systems. Nevertheless, generally the
interactions between plasma particles and the quantum nature of free
electrons cannot be neglected. As a result, the general ionization
equilibrium equations can be reformed as follows,
\begin{equation}
\frac{n_{i}}{n_{i-1}}=\frac{U_{i}}{U_{i-1}}\exp\left[-\xi-I_{i}^{\text{eff}}/k_{B}T\right],\qquad(i=1,...Z_{\text{max}})\label{eq:GSH}
\end{equation}
which are also called generalized Saha equations. In comparison with
Saha equation, the isolated ionization potential $I_{i}$ is replaced
by the effective ionization potential
\begin{eqnarray}
I_{i}^{\text{eff}} & = & I_{i}-\left(\frac{\partial}{\partial
N_{i-1}}-\frac{\partial}{\partial N_{i}}-\frac{\partial}{\partial
N_{e}}\right)F^{\text{non-id}},
\end{eqnarray}
which includes the chemical potential shift contributed by the
interacting part of the free energy and is known as pressure-induced
ionization potential lowering. Obviously, the plasma composition is
determined by the effective ionization potentials, which are in turn
determined by the interacting part of the free energy function.
Therefore, the construction of the free energy function is of
fundamental importance.

\subsection{The free energy function}

Then we shall present the chemical model used in this paper, by
specifying the free energy function, which can be decomposed into
the kinetic term, the ionization energy term including internal
excitations, the Coulomb term, the configurational term, etc.

\subsubsection{The kinetic free energy}

The kinetic term, which is contributed by the translational motion
of heavy particles (He, $\text{He}^{+}$, $\text{He}^{2+}$) with
Maxwell-Boltzmann statistics and that of free electrons with
Fermi-Dirac statistics, can be written as

\begin{eqnarray}
F_{\text{id}} & = & \sum_{i=0}^{2}N_{i}k_{B}T\left[\ln\left(\frac{N_{i}}{V}\left(\frac{2\pi\hbar^{2}}{m_{i}k_{B}T}\right)^{3/2}\right)-1\right]\nonumber \\
 &  & +N_{e}k_{B}T\left(\xi-\frac{2}{3}\frac{I_{3/2}(\xi)}{I_{1/2}(\xi)}\right),
\end{eqnarray}
Here, $N_{0}$, $N_{1}$, $N_{2}$ and $N_{e}$ are the numbers of He,
$\text{He}^{+}$, $\text{He}^{2+}$ and $e$ respectively, $m_{i}$ is
the mass of the corresponding heavy particle, $m_{2}\simeq
m_{1}\simeq m_{0}=m_{\text{He}}$, and the Fermi integrals are
defined by
\begin{equation}
I_{\nu}(\xi)\equiv\int_{0}^{\infty}dx\frac{x^{\nu}}{e^{x-\xi}-1},\qquad(\nu=1/2,\,3/2).
\end{equation}
Note that the free part of the electronic chemical potential
$\xi\equiv\mu_{e}^{\text{id}}/k_{B}T$ is determined by the number
density of free electrons,

\begin{equation}
I_{1/2}(\xi)=\frac{\sqrt{\pi}}{4}\frac{N_{e}}{V}\left(\frac{2\pi\hbar^{2}}{m_{e}k_{B}T}\right)^{3/2},
\end{equation}
where the electronic spin has been taken into account.

\subsubsection{The ionization free energy}

The ionization energy term is given by

\begin{eqnarray}
F_{\text{i-ex}} & = &
-N_{1}k_{B}T\ln2+N_{1}I_{1}+N_{2}(I_{1}+I_{2}),
\end{eqnarray}
where $I_{1}=24.6$ eV, $I_{2}=54.4$ eV. Note that the bound states
of He and $\text{He}^{+}$ are included in the simplest way by
ignoring all excited states, which is a good approximation when the
temperature is much lower than the first excited level, since in
this paper we are constrained to low temperatures when pressure
effects dominate over temperature effects.

\subsubsection{The Coulomb free energy}

There remains no exact formula to calculate the Coulomb free energy
of charged particles (free electrons, bare ions and ions with bound
electrons). In the SCvH model \cite{SCvH1995}, interactions between
the charged particles, including $\text{He}^{2+}$, $\text{He}^{+}$
and e, are described by the Debye-H\"uckel approximation, which is
constrained in the weak coupling regime. In Ref.\cite{WC2005}, the
Coulomb contributions are estimated in the framework of ion sphere
model. In Refs.\cite{JSR2003,Schwarz2005,Chen2007}, the Pad\'e
interpolation formula, which have been developed for fully ionized
plasma \cite{Stoltzmann1996-2000,CP1998}, are used to describe the
Coulomb interactions between charged particles in partially ionized
plasma over a wide range of densities and temperatures. In this
paper, we shall try the Pad\'e interpolation formula for the
approximate description of Coulomb interactions. The Coulomb term
can be split into four parts as follows,

\begin{equation}
F_{\text{coul}}=F_{\text{ee}}^{\text{x}}+F_{\text{ee}}^{\text{c}}+F_{\text{ii}}^{\text{c}}+F_{\text{ie}}^{\text{c}},\label{eq:FCoul}
\end{equation}
where x and c denote the exchange and the correlation terms
respectively, $F_{\text{ee}}$ and $F_{\text{ii}}$ correspond to the
electron and the ion fluid, $F_{\text{ie}}$ describes the effect of
the ion-electron interaction. The detailed formula of the four parts
can be found in Ref.\cite{Stoltzmann1996-2000}.

\subsubsection{The configurational free energy}

The $\text{He}^{+}$ ion, which carries a net charge and a bound
electron, has both the charge nature and the finite size nature, of
which the former is considered in the Coulomb term, and the latter
should be considered additionally. At low densities when the
particles are far enough apart so that they feel only the Coulomb
part of the $\text{He}^{+}$ potential but not the very short range
effect of the bound electron, the finite size of $\text{He}^{+}$can
be neglected \cite{SCvH1995}. At high densities, the $\text{He}^{+}$
ions can be modeled by hard spheres with the diameter estimated as
twice the expectation value of the radius $r$ in the unperturbed
hydrogen-like 1s-state, $d_{1}=1.5\, a_{\text{B}}$, where
$a_{\text{B}}$ is the Bohr radius \cite{FKE1992}. The diameter of
the bare nuclei $\text{He}^{2+}$ is neglected.

The interactions in the subsystem of neutral atoms are described by
the hard-sphere variational formulation of fluid perturbation theory
using the effective pair-wise additive potential, which takes the
form of Aziz and Slaman \cite{Aziz} for $r\geq1.8$ \AA, and that of
Ceperley and Partridge \cite{CP} for $r<1.8$ \AA. To mimic the
softening due to many-body effects at high density, this effective
pair potential is modified by a density-dependent function
\cite{WC2005},

\begin{equation}
\tilde{\Phi}_{\text{eff}}(r)=\left(1-C+\frac{C}{1+D\rho}\right)\Phi_{\text{eff}}(r),
\end{equation}
where the two parameters $(C,D)=(0.44,0.8\,\text{cm}^{3}\text{/g})$
are optimized to reproduce the experimental measures of adiabatic
sound velocity. Note that the fluid perturbation theory of the
neutral subsystem is disturbed by the presence of the finite-sized
$\text{He}^{+}$ ions. The mixture of the $\text{He}^{+}$ hard
spheres and the hard core part of the He atoms is described by the
Mansoori formula \cite{Mansoori1971},
\begin{eqnarray}
F_{\text{hc}}/Nk_{B}T & = & -\frac{3}{2}(1-y_{1}+y_{2}+y_{3})+\frac{3y_{2}+2y_{3}}{1-\eta}\nonumber \\
 &  & +\frac{3}{2}\frac{1-y_{1}-y_{2}-y_{3}/3}{(1-\eta)^{2}}+(y_{3}-1)\ln(1-\eta),\nonumber \\
\end{eqnarray}
where
\begin{eqnarray}
y_{1} & = & \Delta_{01}\left(\sqrt{\frac{d_{1}}{d_{0}}}+\sqrt{\frac{d_{0}}{d_{1}}}\right),\nonumber \\
y_{2} & = & \Delta_{01}\left(\frac{\eta_{0}}{\eta}\sqrt{\frac{d_{1}}{d_{0}}}+\frac{\eta_{1}}{\eta}\sqrt{\frac{d_{0}}{d_{1}}}\right),\nonumber \\
y_{3} & = & \left[\left(\frac{\eta_{0}}{\eta}\right)^{2/3}x_{0}^{1/3}+\left(\frac{\eta_{1}}{\eta}\right)^{2/3}x_{1}^{1/3}\right]^{3},\nonumber \\
N & = & N_{0}+N_{1}+N_{2},\quad x_{0}=N_{0}/N,\quad x_{1}=N_{1}/N,\nonumber \\
\eta & = & \eta_{0}+\eta_{1},\quad\eta_{0}=\frac{1}{6}\pi d_{0}^{3}\frac{N_{0}}{V},\quad\eta_{1}=\frac{1}{6}\pi d_{1}^{3}\frac{N_{1}}{V},\nonumber \\
\Delta_{01} & = & \frac{\sqrt{\eta_{0}\eta_{1}}}{\eta}\frac{\left(d_{0}-d_{1}\right)^{2}}{d_{0}d_{1}}\sqrt{x_{0}x_{1}},\nonumber \\
\end{eqnarray}
with the $\text{He}^{+}$ hard sphere diameter
$d_{1}=1.5a_{\text{B}}$ and the He hard core diameter $d_{0}$ to be
determined.

The perturbation part of the He-He interaction is given by

\begin{eqnarray}
F_{\text{pert}} & = & \frac{2\pi
N_{0}^{2}}{V}\int_{d_{0}}^{\infty}g_{\text{hs}}(r,\tilde{\eta}_{0})\tilde{\Phi}_{\text{eff}}(r)r^{2}dr,
\end{eqnarray}
where $g_{\text{hs}}(r,\tilde{\eta}_{0})$ is the radial distribution
function in a system of hard spheres with the packing fraction
$\tilde{\eta}_{0}$. Note that the packing fraction
$\tilde{\eta}_{0}$ means the ratio of the space occupied by the core
part of the He atoms to the total space allowed for the He atoms.
However, due to the presence of the $\text{He}^{+}$ ions, the total
space allowed for the He atoms is decreased to be $V(1-\eta_{1})$,
and consequently, the packing fraction can be estimated as
$\tilde{\eta}_{0}=\eta_{0}/(1-\eta_{1})$.

The He hard core diameter $d_{0}$ is determined by minimizing the
configurational free energy
$F_{\text{conf}}=F_{\text{hc}}+F_{\text{pert}}$.

\subsubsection{The quantum correction of inter-atomic interaction}

Note that when the de Broglie wavelength of He atoms becomes
comparable to the range of the inter-atomic potential, the He atoms
no longer interact as classical point-like particles, and
consequently, the quantum effects should be included, which in the
Wigner-Kirkwood expansion to the first non-vanishing term
\cite{AC1994,SCvH1995,WC2005,Chen2007} can be written as

\begin{equation}
F_{\text{qm}}=\frac{\hbar^{2}N_{0}^{2}}{24m_{0}k_{B}TV}\int_{d_{0}}^{\infty}\nabla^{2}\tilde{\Phi}_{\text{eff}}(r)g_{\text{hs}}(r,\tilde{\eta}_{0})r^{2}dr.
\end{equation}
The strength of the quantum correction can be estimated by the
parameter$\lambda_{\text{DB}}/r_{\text{m}}$, where
$\lambda_{\text{DB}}$ is the de Broglie wavelength of He atoms and
$r_{\text{m}}$ is the radial distance of the inter-atomic potential
minimum.

\subsection{The plasma composition and the thermodynamical quantities}

By minimizing the total free energy function
\begin{eqnarray}
F & = &
F_{\text{id}}+F_{\text{i-ex}}+F_{\text{coul}}+F_{\text{conf}}+F_{\text{qm}},
\end{eqnarray}
with respect to the abundances of the ionic species
$\text{He}^{+}$and $\text{He}^{2+}$, we obtain the free energy and
the plasma composition simultaneously. Then the thermodynamical
quantities such as pressure, internal energy, and entropy can be
derived from the free energy by thermodynamic relations. Note that
minimizing the free energy function to obtain the plasma composition
is equivalent to solving the generalized Saha equations.

\section{Results and Discussions}

The studies will proceed in two steps. First, we are to investigate
the thermodynamical instability induced by pressure ionization,
based on the model described above. Then, we are to examine the
construction of the free energy function and achieve good agreement
with the first principles simulations
\cite{Kietzmann2007,Militzer2009,Militzer2006} by weakening the
Coulomb term.

\begin{figure}
\includegraphics[bb=0 0 700 500, width=8cm]{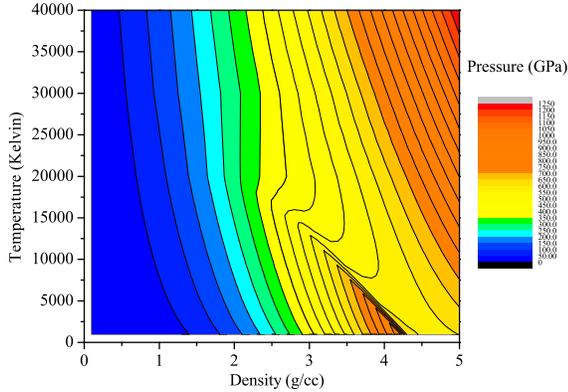}
\caption{\label{fig:Press_Contour} (Color online) The pressure
contour plot of fluid helium as a function of temperature and
density. A region of van der Waals-like loops can be observed.}
\end{figure}

\begin{figure}
\includegraphics[bb=0 0 800 550, width=8cm]{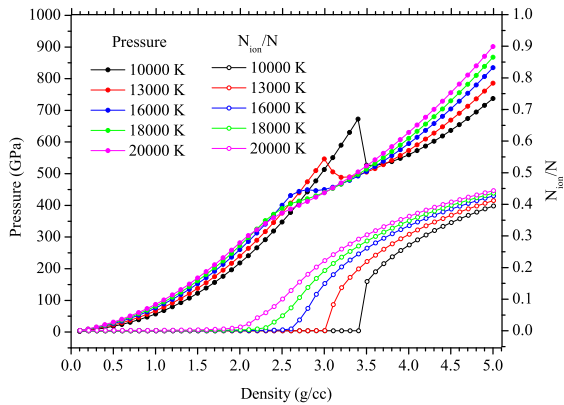}
\caption{\label{fig:isotherms_Annealing} (Color online) The pressure
and ionization degree of helium along various isotherms. The close
connection between the behavior of the pressure curve and the
density dependence of ionization degree is shown. }
\end{figure}

\begin{figure}
\includegraphics[bb=0 0 750 500, width=8cm]{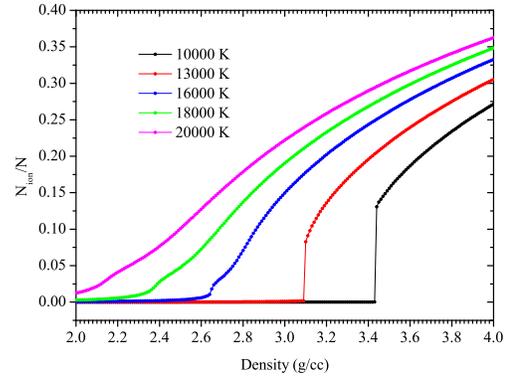}
\caption{\label{fig:isotherms_Annealing_dense} (Color online) The
ionization degrees of helium along the isotherms are plotted at
densely spaced density points. The curve of 10000 K is discontinuous
between 3.43 g/cc and 3.44 g/cc. The curve of 13000 K is
discontinuous between 3.09 g/cc and 3.10 g/cc. The curve of 16000 K
is continuous, although it is extraordinarily steep between 2.64
g/cc and 2.66 g/cc. }
\end{figure}

\begin{figure}
\includegraphics[bb=0 50 800 550, width=8cm]{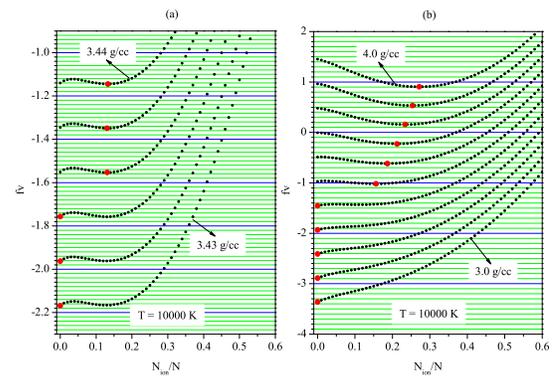}
\caption{\label{fig:scan_10kK} (Color online) The curves selected
from the free energy surfaces over the ionic abundances.
$N_{1}=N_{\text{ion}}$, $N_{2}=0$, $f_{v}=F/Nk_{B}T$. Note that the
plotted values of $f_{v}$ include density-dependent shifts in order
to make the curves more distinguishable. (a) The densities are
equally spaced from 3.430 g/cc to 3.440 g/cc. (b) The densities are
equally spaced from 3.0 g/cc to 4.0 g/cc. The red circle on each
curve denotes its global minimum. The grid lines are shown to guide
the eyes. }
\end{figure}

\begin{figure}
\includegraphics[bb=0 70 800 550, width=8cm]{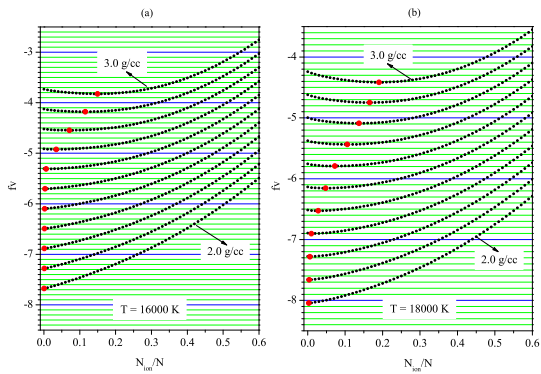}
\caption{\label{fig:scan_16kK_18kK} (Color online) The curves
selected from the free energy surfaces over the ionic abundances.
$N_{1}=N_{\text{ion}}$, $N_{2}=0$, $f_{v}=F/Nk_{B}T$. Note that the
plotted values of $f_{v}$ include density-dependent shifts in order
to make the curves more distinguishable. (a) T=16000 K. (b) T =
18000 K. The densities are equally spaced from 2.0 g/cc to 3.0 g/cc.
The red circle on each curve denotes its global minimum. The grid
lines are shown to guide the eyes.}
\end{figure}

\begin{figure}
\includegraphics[bb=0 0 750 550, width=8cm]{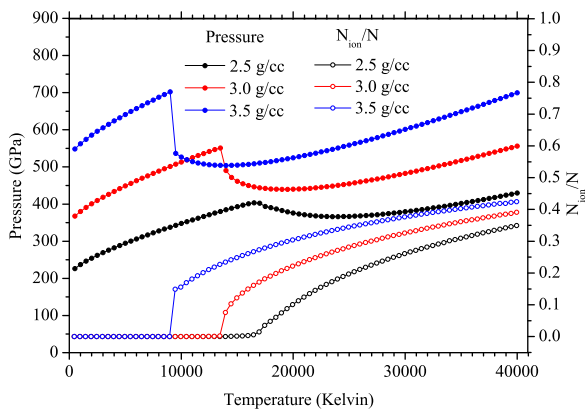}
\caption{\label{fig:isochores_Annealing} (Color online) The pressure
and ionization degree of helium along various isochores at the
densities of 2.5 g/cc, 3.0 g/cc and 3.5 g/cc. The close connection
between the behavior of the pressure curve and the temperature
dependence of ionization degree is shown. }
\end{figure}

\begin{figure}
\includegraphics[bb=0 0 700 500, width=8cm]{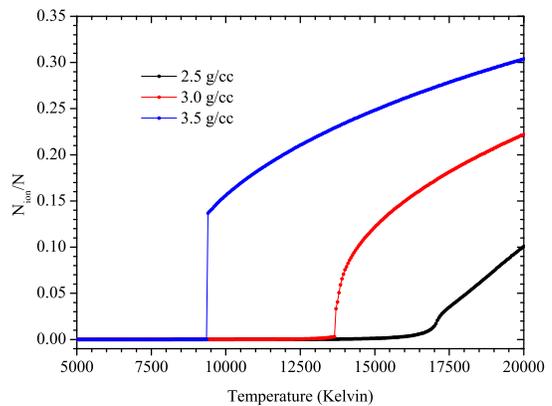}
\caption{\label{fig:isochores_Annealing_dense} (Color online) The
ionization degrees of helium along the isochores are plotted at
densely spaced temperature points. The curve of 3.5 g/cc is
drastically discontinuous between 9350 K and 9400 K. The curve of
3.0 g/cc is a bit discontinuous between 13300 K and 13350 K. The
curve of 2.5 g/cc is continuous, although it is extraordinarily
steep between 17000 K and 17250 K. }
\end{figure}

\begin{figure}
\includegraphics[bb=0 50 750 520, width=8cm]{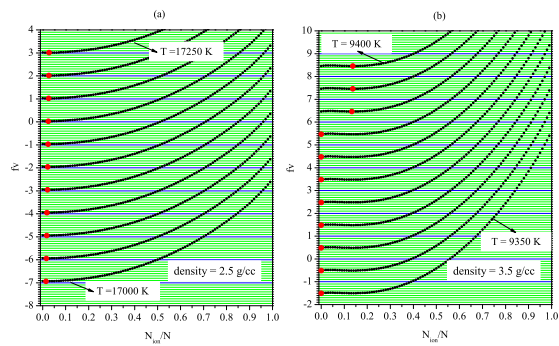}
\caption{\label{fig:scan_2.5gcc_3.5gcc} (Color online) The curves
selected from the free energy surfaces over the ionic abundances.
$N_{1}=N_{\text{ion}}$, $N_{2}=0$, $f_{v}=F/Nk_{B}T$. Note that the
values of $f_{v}$ shown include temperature-dependent shifts in
order to make the curves more distinguishable. (a) density = 2.5
g/cc. The temperatures are equally spaced from 17000 K to 17250 K.
(b) density = 3.5 g/cc. The temperatures are equally spaced from
9350 K to 9400 K. The red circle on each curve denotes its global
minimum. The grid lines are shown to guide the eyes. }
\end{figure}

\subsection{The thermodynamical instability induced by pressure ionization}

First we present the contour plot of the pressure calculated by the
model described above, in a density region between 0.01 g/cc and 5.0
g/cc and for temperatures from 1000 K to 50000 K. As shown in
Fig.\ref{fig:Press_Contour}, a non-trivil region of van der
Waals-like loops can be observed, which may be connected with
pressure ionization. In order to study the details of van der
Waals-like loops and how they are related to pressure ionization, we
have plotted the pressure as well as the ionization degree along
several selected isotherms crossing the region of van der Waals-like
loops. In order to study the temperature dependence of pressure
ionization, we have plotted the pressure as well as the ionization
degree along several selected isochores crossing the region of van
der Waals-like loops.

It is important to highlight that the thermodynamical instability is
generally located in the partial ionization regime of low
temperatures and high densities when pressure effects dominate over
temperature effects, because it is pressure ionization that causes
the thermodynamical instability. Actually, by solving the Saha
equations without the pressure-induced ionization potential lowering
one can easily verify that the temperature ionization always occurs
smoothly and induces no thermodynamical instability.

\subsubsection{The isotherms}

The isotherms for selected temperatures crossing the region of van
der Waals-like loops are plotted in
Fig.\ref{fig:isotherms_Annealing}. It is seen that for 10000 K and
13000 K the pressure drops discontinuously at the top point of the
van der Waals-like loop when the ionization degree jumps to a finite
value abruptly, and for 16000 K the pressure remains smooth when
ionization occurs continuously. As the temperature rises, the van
der Waals-like loop becomes smaller and finally disappears. At
higher temperatures of 18000 K and 20000 K, the van der Waals-like
loop has disappeared, but the slope of the pressure isotherm is
still bent over, and as a result, the pressure of 18000 K is higher
than that of 20000 K, which is opposite to the regions of both low
densities and high densities. Note that the similar phenomenon of
the coexistence pressure decreasing with the temperature was also
found in hydrogen dissociation/ionization \cite{Beule1999PRB}. Above
a critical temperature of about 20000 K, the thermodynamical
instability disappears due to temperature effects.

In Fig.\ref{fig:isotherms_Annealing_dense} by plotting the
ionization degrees at densely spaced density points, it is further
confirmed that the pressure ionization at sufficiently low
temperatures occurs discontinuously and that at higher temperatures
occurs continuously, which has yet to be explained. Note that the
plasma composition is obtained by minimizing the free energy
function with respect to the abundances of the ionic species
$\text{He}^{+}$and $\text{He}^{2+}$, so the straightforward way to
explain the behavior of the ionization degree is by investigating
the free energy surfaces over the abundances of ionic species, of
which the minimum location corresponds to the plasma composition. We
have examined the evolvement of the free energy surface at the
temperature of 10000 K with the increasing density, by inspecting it
at equally spaced densities between the discontinuous points in
Fig.\ref{fig:isotherms_Annealing_dense}. Since throughout the
evolvement its minimum location is always on the line corresponding
to $N_{2}/N=0$, and hence, we only have to plot the free energy
lines as a function of $N_{1}/N$. From Fig.\ref{fig:scan_10kK} it is
found that the discontinuous rise of the ionization degree is due to
the emergence of another local minimum which later becomes the
global minimum abruptly. A similar examination is performed for the
temperatures of 16000 K and 18000 K in Fig.\ref{fig:scan_16kK_18kK},
which shows a single minimum shifting continuously.

From Figs.\ref{fig:isotherms_Annealing} and
\ref{fig:isotherms_Annealing_dense} the behaviors of pressure
ionization and its induced thermodynamical instability along the
isotherms can be summarized as follows: along the isotherms of
sufficiently low temperatures, the pressure ionization occurs
discontinuously, which simultaneously induces a discontinous drop of
the pressure; along the isotherms of a little higher temperatures,
the pressure ionization occurs continuously, which simultaneously
induces a smooth van der Waals-like loop of the pressure; along the
isotherms of even higher temperatures, the smooth van der Waals-like
loop becomes smaller and finally disappears to become bending over;
above the critical temperature, pressure ionization no longer causes
thermodynamical instability, due to significant temperature effects.
Note that the behaviors of the pressure-density relations here are
in excellent agreement with those from the first principles
simulations \cite{LHR2010,Morales2010b}.

\subsubsection{The isochores}

The isochores for selected densities crossing the region of van der
Waals-like loops are plotted in Fig.\ref{fig:isochores_Annealing}.
Note that for high densities, there is a drop on the isochore, like
the findings of the first principles simulations
\cite{Bonev2004b,Vorberger2007} on hydrogen/deuterium. While the
first principles simulations \cite{Bonev2004b,Vorberger2007}
disagree whether the pressure isochore is a smooth curve with a
region of negative slope or whether the pressure isochore has a
discontinuous drop, the pressure isochores here demonstrate that
both cases are possible. It is shown that at the density of 2.5
g/cc, the pressure isochore appears as a smooth curve with a
negative slope region due to the continuous rise of the ionization
degree, but at the higher densities of 3.0 g/cc and 3.5 g/cc, the
pressure exhibits a discontinuous drop due to the discontinuous rise
of the ionization degree. The behavior of the ionization degree is
further confirmed by the plot at densely spaced temperature points
in Fig.\ref{fig:isochores_Annealing_dense}. The evolvement of the
free energy minimum with the rising temperature is shown in
Fig.\ref{fig:scan_2.5gcc_3.5gcc}. Since throughout the evolvement
the minimum of the free energy surface is always located on the line
corresponding to $N_{2}/N=0$, and hence, we only have to plot the
free energy lines as a function of $N_{1}/N$. It is observed that at
the density of 2.5 g/cc the single minimum shifts continuously with
the rising temperature, while at the density of 3.5 g/cc another
local minimum appears and gets lower with the rising temperature, so
that at a certain temperature the global minimum shifts
discontinuously from the local minimum of zero to that of a finite
value.

From Figs.\ref{fig:isochores_Annealing} and
\ref{fig:isochores_Annealing_dense} the behaviors of pressure
ionization and its induced thermodynamical instability along the
isochores can be summarized as follows: along the isochores of
sufficiently high densities, the pressure ionization occurs
discontinuously, which simultaneously induces a discontinous drop of
the pressure; along the isochores of a little lower densities, the
pressure ionization occurs continuously, which simultaneously
induces a continuous fall and rise of the pressure. Note that the
behaviors of the pressure-temperature relations here are in
excellent agreement with those from the first principles simulations
\cite{Bonev2004b,Vorberger2007}.

\subsection{The examination of the free energy function}

In the preceding subsection, we have shown the thermodynamical
instability induced by pressure ionization based on the chemical
model described in Sec.II. Nonetheless, the first principles
simulations such as Refs. \cite{Kietzmann2007,Militzer2009} did not
find any instability associated with pressure ionization in fluid
helium. As is known, the plasma composition and the thermodynamic
properties are determined by the free energy function, of which,
therefore, a bit of uncertainty may bring changes to the pressure
ionization phenomena. Note that the Coulomb free energy is described
by the Pad\'e interpolation formula \cite{Stoltzmann1996-2000}
developped for fully ionized electron-ion plasma. Since there is
currently no exact formula developped for describing the Coulomb
energy among charged particles in partially ionized plasma, which is
also beyond the scope of this paper, we have tried to tune the
Coulomb term in a crude way and found that it achieves good
agreement with the first principles simulations when the Coulomb
term in Eq.\eqref{eq:FCoul} is weakened by a factor of
$e^{-\Gamma_{\text{ion}}^{0.5}-\Gamma_{\text{e}}^{0.5}}$, with the
Coulomb coupling parameters $\Gamma_{\text{ion}}$and $\Gamma_{e}$
defined in Ref.\cite{Stoltzmann1996-2000}.

\subsubsection{No thermodynamical instability}

\begin{figure}
\includegraphics[bb=0 0 750 500, width=8cm]{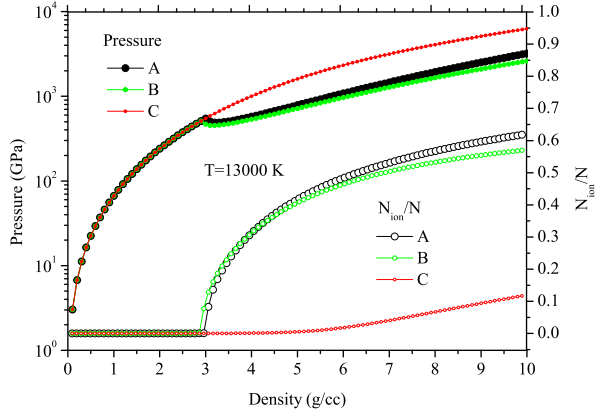}
\caption{\label{fig:isotherms_variation} (Color online) The pressure
and ionization degree of helium along the isotherm at the
temperature of 13000 K, calculated by the chemical model and its
variants. A: the original model. B: the model with
$\tilde{\eta}_{0}=\eta_{0}$. C: the model with the weakened Coulomb
term.}
\end{figure}

\begin{figure}
\includegraphics[bb=0 0 750 550, width=8cm]{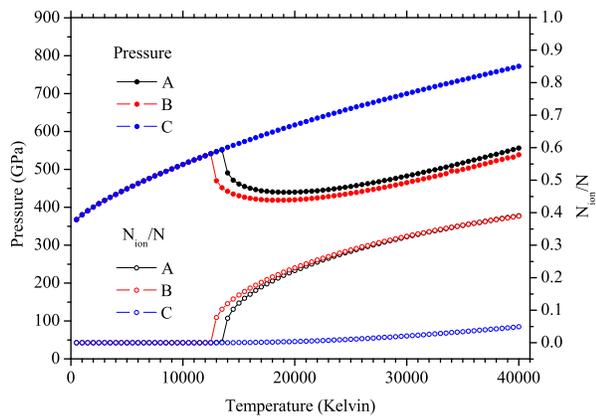}
\caption{\label{fig:isochores_variations} The pressure and
ionization degree of helium along the isochore at the density of 3.0
g/cc, calculated by the chemical model and its variants. The
meanings of {}``A, B, C'' are the same as in
Fig.\ref{fig:isotherms_variation}.}
\end{figure}

Using the weakened Coulomb term, the studies on pressure ionization
are repeated and the results are compared with the original model in
Fig.\ref{fig:isotherms_variation} and
Fig.\ref{fig:isochores_variations}. Note that the pressure
ionization becomes strongly suppressed and brings no thermodynamical
instability. In addition, the effect of the finite size of
$\text{He}^{+}$ ions on the He fluid perturbation theory is also
investigated, by repeating the calculations using the packing
fraction of $\tilde{\eta}_{0}=\eta_{0}$. It can be observed that the
difference between the result of $\tilde{\eta}_{0}=\eta_{0}$ and
that of $\tilde{\eta}_{0}=\eta_{0}/(1-\eta_{1})$ is relatively small
and becomes more apparent with the increasing density.

\subsubsection{The EOS in good agreement with the first principles simulations}

The EOS models are usually checked by Hugoniot data. The
double-shock Hugoniot curves are calculated by the original model,
the model with $\tilde{\eta}_{0}=\eta_{0}$ and the model with the
weakened Coulomb term, using the initial states in accordance with
the Nellis \textit{et al.} experiment \cite{Nellis1984}.

\begin{figure}
\includegraphics[bb=0 0 750 500, width=8cm]{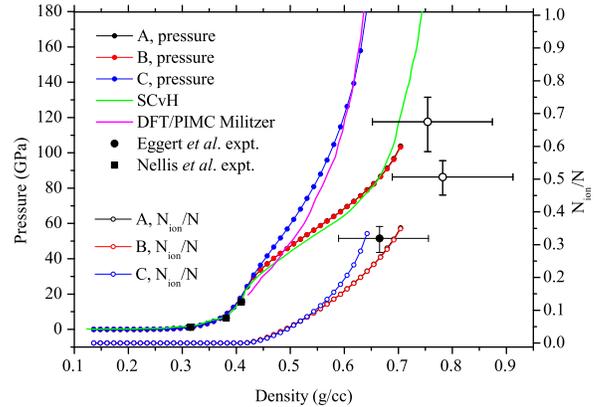}
\caption{\label{fig:Hug1st} (Color online) The calculated Hugoniot
data of fluid helium from (0.1235 g/cc, 4.3 K), including the
pressure and the ionization degree. The meanings of {}``A, B, C''
are the same as in Fig.\ref{fig:isotherms_variation}. Also shown are
the first-shock data from Nellis \textit{et al.} \cite{Nellis1984}
(black square symbols), the shock data from Eggert \textit{et al.}
\cite{Eggert2008} (circle symbols), the model calculation from SCvH
\cite{SCvH1995} (green line) and the DFT/PIMC simulation from
Ref.\cite{Militzer2006} (magenta line). }
\end{figure}

\begin{figure}
\includegraphics[bb=0 0 700 500, width=8cm]{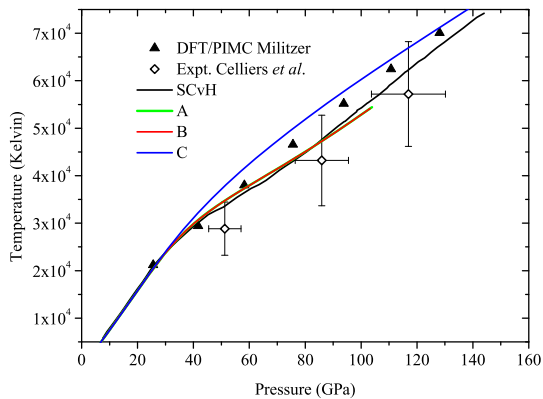}
\caption{\label{fig:HugT1st} (Color online) The calculated Hugoniot
temperature of fluid helium from (0.1235 g/cc, 4.3 K). The meanings
of {}``A, B, C'' are the same as in
Fig.\ref{fig:isotherms_variation}. Also shown are the shock data
from Celliers \textit{et al.} \cite{Celliers2010} (open diamond
symbols), the model calculation from SCvH \cite{SCvH1995} (black
line) and the DFT/PIMC simulation from Ref.\cite{Militzer2006}
(black triangles). }
\end{figure}

The first-shock curves are shown in Fig.\ref{fig:Hug1st}, in
comparison with experimental data \cite{Nellis1984,Eggert2008} and
other theoretical results \cite{SCvH1995,Militzer2006}. Note that
the calculated Hugoniot curves begin to diverge when ionization
occurs. The original model is in good agreement with the SCvH model
\cite{SCvH1995}, but much softer than the first principles
simulation \cite{Militzer2006}. By using $\tilde{\eta}_{0}=\eta_{0}$
the results are not much changed. By using the weakened Coulomb
term, the calculated Hugoniot pressure is much enhanced to achieve
good agreement with the first principles simulation
\cite{Militzer2006}. As for the experimental data, the Nellis
\textit{et al.} shock data \cite{Nellis1984} has not reached the
ionization regime, which thus cannot be employed to distinguish the
chemical models; the Eggert \textit{et al.} shock data
\cite{Eggert2008} has indeed probed into the ionization regime, but
its Hugoniot pressure is much lower than all the theoretical
results. In Fig.\ref{fig:HugT1st} the Hugoniot temperature is
plotted versus the Hugoniot pressure. Note that the original model
is in good agreement with the SCvH model \cite{SCvH1995}, and the
model with the weakened Coulomb term is slightly above the first
principles simulation \cite{Militzer2006}. The experimental data
\cite{Celliers2010} is relatively low, with the error bars covering
some theoretical results.

\begin{figure}
\includegraphics[bb=0 0 750 510, width=8cm]{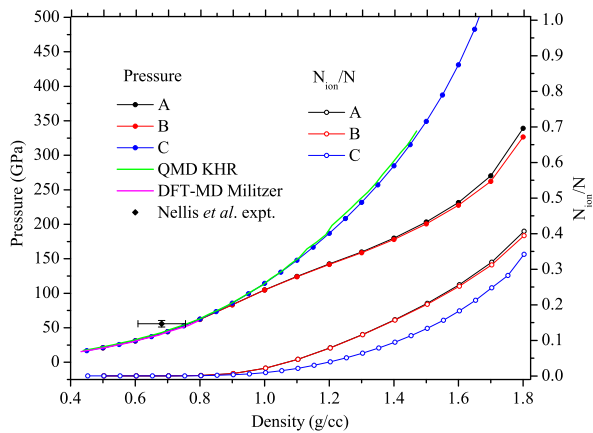}
\caption{\label{fig:Hug2nd} (Color online) The calculated Hugoniot
data of fluid helium from (9.79 cc/mol,13.8 GPa), including the
pressure and the ionization degree. The meanings of {}``A, B, C''
are the same as in Fig.\ref{fig:isotherms_variation}. Also shown are
the seond-shock data from Nellis \textit{et al.} \cite{Nellis1984}
(black diamond symbols) and the first principles simulation results
of QMD KHR \cite{Kietzmann2007} and DFT-MD Militzer
\cite{Militzer2006}.}
\end{figure}

The second-shock curves are shown in Fig.\ref{fig:Hug2nd}, in
comparison with experimental data \cite{Nellis1984} and first
principles simulations \cite{Kietzmann2007,Militzer2006}. It is
further confirmed that by using the weakened Coulomb term, the
calculated Hugoniot pressure can be substantially enhanced and
achieve good agreement with the first principles simulations
\cite{Kietzmann2007,Militzer2006}. Note that the shock data of
{}``QMD KHR'' \cite{Kietzmann2007} is from a slightly different
initial state.

\begin{figure}
\includegraphics[bb=50 0 750 500, width=8cm]{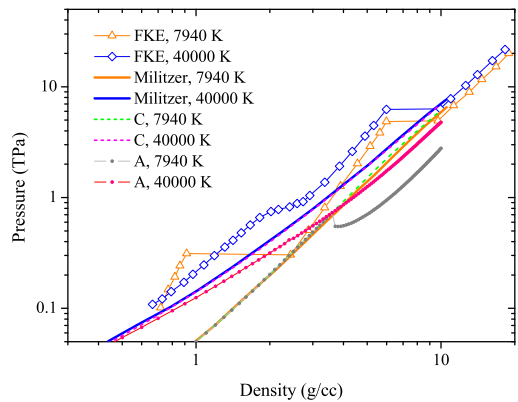}
\caption{\label{fig:Pressure} (Color online) The pressure isotherms
calculated by the original model (A) and the model with the weakened
Coulomb term (C) are shown, in comparison with the FKE chemical
model (FKE) \cite{FKE1992} and the first principles simulations
(Militzer) \cite{Militzer2009}. }
\end{figure}

\begin{figure}
\includegraphics[bb=0 0 700 500, width=8cm]{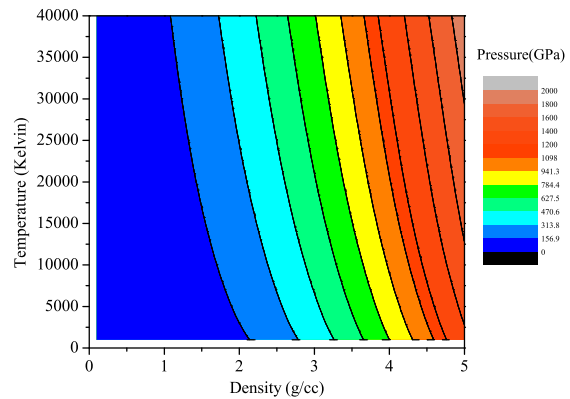}
\caption{\label{fig:EOS} (Color online) The pressure contour plot of
fluid helium, calculated from the model with the weakened Coulomb
term. }
\end{figure}

The EOS models can also be checked by pressure isotherms. It can be
observed from Fig.\ref{fig:Pressure} that the chemical models used
in this paper are in much better agreement with the DFT/PIMC result
\cite{Militzer2009} than the FKE chemical model \cite{FKE1992}. The
model with the weakened Coulomb term achieves particularly good
agreement with the DFT/PIMC result \cite{Militzer2009}. As shown in
Fig.\ref{fig:EOS}, the contour plot of the pressure from the model
with the weakened Coulomb term does not exhibit any instability
region, in contrast to the original model shown in
Fig.\ref{fig:Press_Contour}.

\section{Conclusions and Outlooks}

The pressure ionization has received intensive attention, due to
fundamental interest in condensed matter physics, and also due to
the importance in providing EOS for astrophysics and ICF research.
In the first principles simulations, the pressure ionization of
fluid hydrogen occurs via a first order liquid-liquid phase
transition, and that of fluid helium, in contrast, occurs smoothly
without any indication of first order transition. However, in the
chemical model calculations the pressure ionization of fluid helium
also occurs via a first order transition typically known as plasma
phase transition. In this paper we have carried out a systematic
study of pressure ionization in fluid helium, in the framework of
chemical models. It is demonstrated that when pressure ionization
occurs and whether it induces thermodynamical instability are
dependent on the construction of the free energy function. In the
chemical model described in Sec. II, we have found the
thermodynamical instability, which is induced by pressure ionization
and is manifested by a discontinuous drop or a continuous fall and
rise along the pressure-density and pressure-temperature curves.
When the chemical model is modified by weakening the Coulomb term of
the free energy function using an empirical factor, the pressure
ionization occurs at higher densities and no longer induces
thermodynamical instability. Moreover, the resulting smooth EOS
achieves good agreement with the first principles simulations
\cite{Kietzmann2007,Militzer2009,Militzer2006}. It is interesting
that the thermodynamical instability induced by pressure ionization
in fluid helium from the chemical model here is very much like that
of the first order liquid-liquid phase transition of fluid hydrogen
from the first principles simulations
\cite{Bonev2004b,Vorberger2007,LHR2010,Morales2010b}. This implies
that there may be some feature shared by the first principles
simulations of molecular dissociation and the chemical model
calculations of pressure ionization, which causes thermodynamical
instability in the similar way. We guess that in analogy to the
emergence of another local minimum in the free energy surface in the
chemical models, the first principles simulations may encounter
another solution of the electronic Kohn-Sham equation.

Note that the great discrepancy between the first principles
simulation results and the shock compression data in the regime of
partial ionization remains to be resolved. Also note that the good
agreement with the first principles simulation results is achieved
by weakening the Coulomb term in a crude way, and it remains
desirable to construct an accurate free energy function, especially
the Coulomb term.

\begin{acknowledgments}
We are grateful to Qi-Li Zhang and Guang-Cai Zhang for helpful
discussions. This work was supported by the National Science
Foundation of China under Grants No.10804011 and No.11204015.
\end{acknowledgments}

\end{CJK}

\end{document}